\documentclass[prl,twocolumn,showpacs,floatfix,amsfonts]{revtex4}
\usepackage{graphicx,graphics,color,epsfig}
\usepackage{bm}
\usepackage{amsmath}
\usepackage{amssymb}
\begin{document}
\preprint{}
\title{
Incommensurate Magnetism around Vortices and Impurities in
High-$T_c$ Superconductors}
\author{Jian-Xin Zhu}
\affiliation{Theoretical Division, Los Alamos National Laboratory,
Los Alamos, New Mexico 87545}
\author{Ivar Martin}
\affiliation{Theoretical Division, Los Alamos National Laboratory,
Los Alamos, New Mexico 87545}
\author{A. R. Bishop}
\affiliation{Theoretical Division, Los Alamos National Laboratory,
Los Alamos, New Mexico 87545}

\begin{abstract}
By solving self-consistently an effective Hamiltonian including
interactions for both antiferromagnetic spin-density wave (SDW)
and $d$-wave superconducting (DSC) orderings, a comparison study
is made for the local magnetic structure around superconducting
vortices and unitary impurities. To represent the optimally doped
regime of cuprates, the parameter values are chosen such that the
DSC is dominant while the SDW is vanishingly small. We show that
when vortices are introduced into the superconductor, an
oscillating SDW is induced around them. The oscillation period of
the SDW is microscopically found, consistent with experiments, to
be eight lattice constants ($8a_0$). The associated charge-density
wave (CDW) oscillates with a period of one half ($4a_0$) of the
SDW. In the case of unitary impurities, we find a SDW modulation
with identical periodicity, however without an associated CDW. We
propose neutron scattering experiments to test this prediction.
\end{abstract}

\maketitle

The common feature of all high-transition-temperature (high-$T_c$)
cuprates is the proximity between antiferromagnetic (AF) and
d-wave superconducting (DSC) phases controlled by the doping---At
low temperatures, the AF order of the parent compounds
 is replaced by a DSC order upon doping. The main theme of
high-$T_c$ superconducivity research centers on how to establish
the connection between these two kinds of orderings, i.e., whether
they exclude each other, or coexist microscopically. In one of the
strategies, the interplay between the AF and DSC ordering is
explored by weakening the superconductivity in the optimally or
slightly overdoped regime. The inelastic neutron scattering (INS)
measurements on both optimally doped and underdoped
La$_{2-x}$Sr$_{x}$CuO$_{4}$ (LSCO) samples by Lake and co-workers
provide the first evidence of the field-induced magnetic
order~\cite{Lake01,Lake02}. The field-induced spin fluctuations
have a spatial periodicity of $8a_{0}$ with the wave vector
pointing along the Cu-O bond directions. The principle
magnetization oscillations in the vortex state are coherent over a
distance $L_{M}>20a_{0}$, substantially longer than the
superconducting coherence length $\xi_0$ ($\sim 5a_{0}$). The
field-induced enhancement of the Bragg peak intensity was also
observed by Khaykovich {\em et al.}~\cite{Khaykovich01} in the
elastic neutron scattering (ENS) measurement on a related
material, La$_{2}$CuO$_{4+y}$ (LCO), but with $L_{M}>100a_{0}$,
indicating the existence of the field-induced static AF order of
$8a_0$ periodicity. Strong AF fluctuations have also been observed
by Mitrovic {\em et al.}~\cite{Mitrovic01} in a high-field nuclear
magnetic resonance (NMR) imaging experiment on near-optimally
doped YBa$_{2}$Cu$_{3}$O$_{7-x}$ (YBCO). More recently, the
scanning tunneling microscopy imaging by Hoffman {\em et
al.}~\cite{Hoffman02} has revealed the quasiparticle states around
the vortex cores in slightly overdoped
Bi$_{2}$Sr$_{2}$CaCu$_{2}$O$_{8+x}$ (BSCCO)  a Cu-O bond-oriented
``checkboard'' pattern with $4a_0$ periodicity. The periodicity
($4a_0$) of charge modulation is one half of that ($8a_0$) of the
field-induced SDW modulation. Theoretically, Demler and
co-workers, by focusing their attention on the far-field regions
outside the vortex core, proposed a phenomenological
model~\cite{Demler01} that when the superconductivity is weakened
by the circulating currents caused by the vortices, a coexisting
SDW plus DSC phase appears surrounding the core; more recently
they also argued~\cite{Zhang01} that the halved periodicity of the
static CDW modulation is associated with the ``Friedel oscillation
of the spin gap''. As an extension of earlier work within the
SO(5) theory~\cite{Zhang97,Arovas97}, originally suggesting an AF
insulating region inside the core, Hu and Zhang~\cite{Hu01} showed
that the vortex-induced AF region can be greater than $\xi_0$, due
to the light effective mass of the dynamic AF fluctuations at
optimal doping. In these two cases, the $8a_0$ periodicity of the
magnetic modulation is not fully understood due to the
phenomenological nature of the models. Distinctly, stripe
models~\cite{Poilblanc89,Zaanen89,Emery90,Zachar98} predict that
the spin modulation of wavelength $\lambda$ in cuprates should be
associated with the charge modulation of wavelength $\lambda/2$.
However, two dimensional modulation is apparent in the STM
image~\cite{Hoffman02}. One of us and C.S. Ting~\cite{Zhu01} have
applied an effective microscopic mean-field model accounting for
the competition between the AF and DSC orderings~\cite{Martin00}
to successfully explain the peak-split structure around the Fermi
surface in the local density of states (LDOS) at the vortex core
center~\cite{Maggio95,Pan00,Hoogenboom01}. Nevertheless, due to
the use of the simplest band structure parameter values, the spin
and charge modulation is too weak for the determination of the
periodicity. In this work, using realistic model parameter values,
we present a comprehensive study of the spin/charge structure
around the vortices and nonmagnetic unitary impurities in
optimally doped high-$T_c$ superconductors. We find, for the first
time, microscopically: (i) For the vortex case, the oscillation
periods of the SDW and the associated CDW are indeed $8a_0$ and
$4a_0$, respectively, in good agreement with the INS~\cite{Lake01}
and STM~\cite{Hoffman02} experiments. (ii) Around the nonmagnetic
unitary impurity, the modulation of the SDW order still exhibits
$8a_0$ periodicity. However, the charge density shows only the
Friedel oscillation with a period of the Fermi wavelength due to
the strong potential scattering from the impurity.

Consider a minimal model defined on a two dimensional (2D) lattice
square , in which the on-site repulsion is solely responsible for
the antiferromagnetism while the nearest neighbor attraction
causes the $d$-wave superconductivity~\cite{Martin00}. With the
application of an external magnetic field and/or in the presence
of nonmagnetic impurities, the effective mean-field Hamiltonian
can be diagonalized by solving self-consistently the Bogoliubov-de
Gennes equation~\cite{Zhu01}:
\begin{equation}
\sum_{j} \left(
\begin{array}{cc}
{\cal H}_{ij,\sigma} & \Delta_{ij}  \\
\Delta_{ij}^{*} & -{\cal H}_{ij,\bar{\sigma}}^{*}
\end{array}
\right) \left(
\begin{array}{c}
u_{j\sigma}^{n} \\ v_{j\bar{\sigma}}^{n}
\end{array}
\right) =E_{n} \left(
\begin{array}{c}
u_{i\sigma}^{n} \\ v_{i\bar{\sigma}}^{n}
\end{array}
\right)  \;. \label{EQ:BdG}
\end{equation}
Here $(u_{i\sigma}^{n},v_{i\bar{\sigma}}^{n})$ is the
quasiparticle wavefunction corresponding to the eigenvalue $E_n$,
the single particle Hamiltonian ${\cal H}_{ij,\sigma}=-t_{ij}
e^{i\varphi_{ij}} +(m_{i\bar{\sigma}}
+\epsilon_{i}-\mu)\delta_{ij}$, where $t_{ij}=t$ for the nearest
neighbor hopping while $t_{ij}=t^{\prime}$ for the next-nearest
neighbor hopping, $\epsilon_i$ is the single site potential
describing the scattering from impurities, and the Peierls phase
factor $\varphi_{ ij}=\frac{\pi}{\Phi_{0}} \int_{{\bf
r}_{j}}^{{\bf r}_{i}} {\bf A}({\bf r})\cdot d{\bf r}$ with
$\Phi_0=hc/2e$ the superconducting flux quantum in the presence of
an externally applied magnetic field. Notice that the
quasiparticle energy is measured with respect to the Fermi energy.
The self-consistency conditions read:
\begin{subequations}
\begin{eqnarray}
m_{i\uparrow}&=&U n_{i\uparrow}=U\sum_{n} \vert
u_{i\uparrow}^{n}\vert^{2} f(E_n)\;,\\
m_{i\downarrow}&=&U n_{i\downarrow}= U\sum_{n}\vert
v_{i\downarrow}^{n}\vert^{2}[1-f(E_n)]\;,
\end{eqnarray}
\end{subequations}
and
\begin{equation}
\Delta_{ij}=\frac{V}{4}\sum_{n} (u_{i\uparrow}^{n}v_{
j\downarrow}^{n*} +v_{i\downarrow}^{n*}u_{j\uparrow}^{n} ) \tanh
\left( \frac{E_{n}}{2k_{B}T}\right)\;,
\end{equation}
where $U,V$ are the strength of the on-site repulsion and the
nearest neighbor attraction, respectively, and the Fermi
distribution function $f(E)=1/[e^{E/k_{B}T}+1]$. Here the
summation is over the eigenstates with both positive and negative
eigenvalues~\cite{Zhu01}. We report results below for two cases at
zero temperature. For the Abrikosov vortex state, the magnetic
field effects enter through the Peierls phase factor
$\varphi_{ij}$ and no impurities are introduced ($\epsilon_i=0$).
For the effects of a single impurity, we set $\mathbf{A}=0$ so
that $\varphi_{ij}=0$. Hereafter we measure the length in units of
the lattice constant $a_0$ and the energy in units of the hopping
integral $t$. To mimic a hole-like Fermi surface, as relevant to
the hole-doped cuprates, we take $t^{\prime}=-0.2$. As a model
calculation, the on-site repulsion and pairing interactions are
taken to be, $U=2.5$ and $V=1.0$. In addition, the filling factor
$n_f=\sum_{i,\sigma} n_{i\sigma} /N_x N_y$ is fixed to be 0.84,
which corresponds to an optimal hole doping $n_h=1-n_{f}=0.16$.
Here $N_x,N_y$ are the linear dimensions of the unit cell under
consideration. We use an exact diagonalization method to solve the
BdG equation~(\ref{EQ:BdG}) self-consistently. In the absence of
magnetic field and impurities, we recover all results reported
in~\cite{Martin00}. In the optimal doping regime ($n_h=0.16$), the
AF SDW order is absent while the DSC order is homogeneous. When
the DSC is weakened by the application of an external magnetic
field or by introducing impurities, the AF SDW will be nucleated
in the region where the DSC order is depressed.

\begin{figure}
\centerline{\psfig{figure=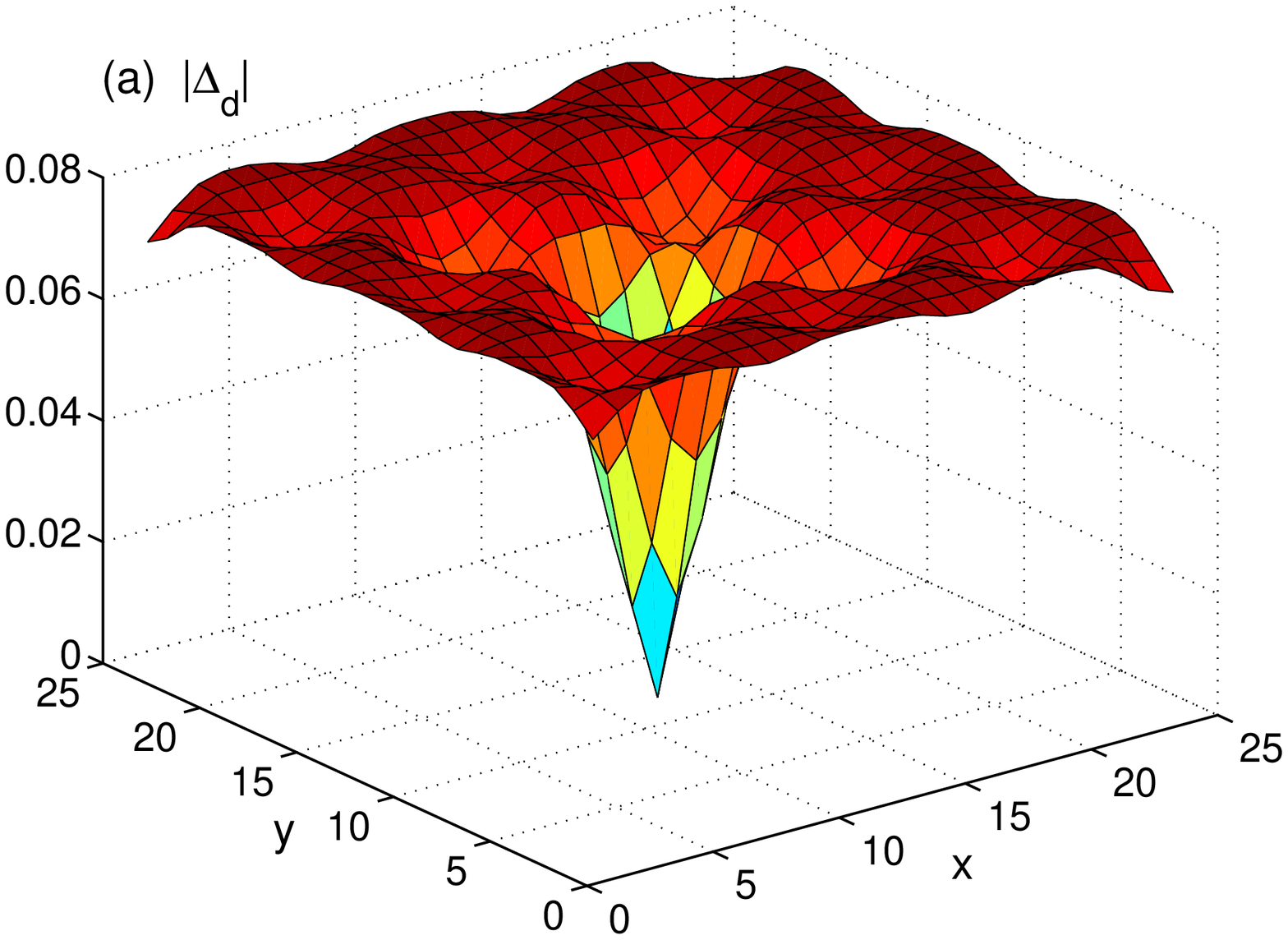,height=3.5cm,width=4.3cm,angle=0}
\psfig{figure=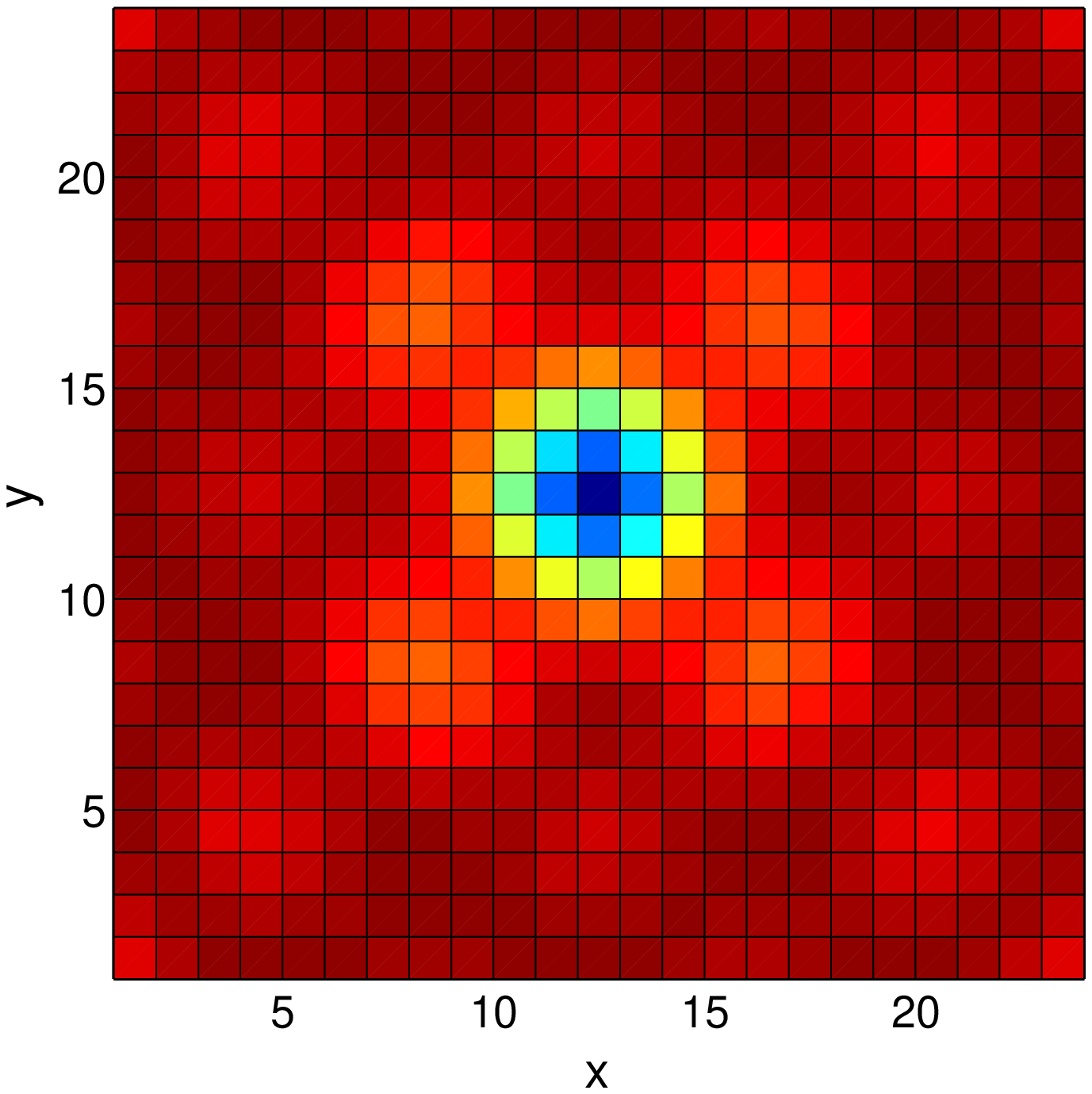,height=3.5cm,width=3.5cm,angle=0}}
\centerline{\psfig{figure=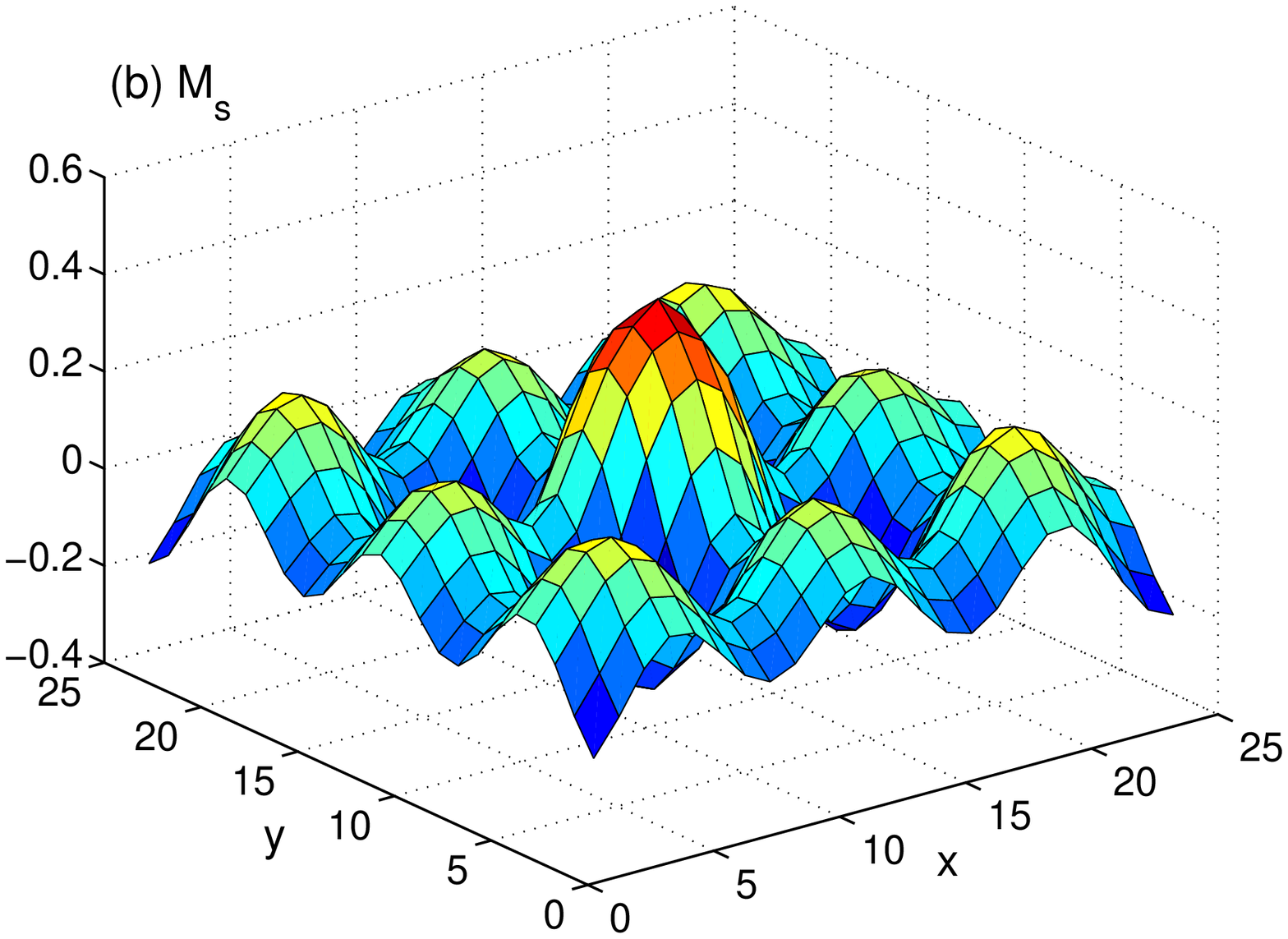,height=3.5cm,width=4.3cm,angle=0}
\psfig{figure=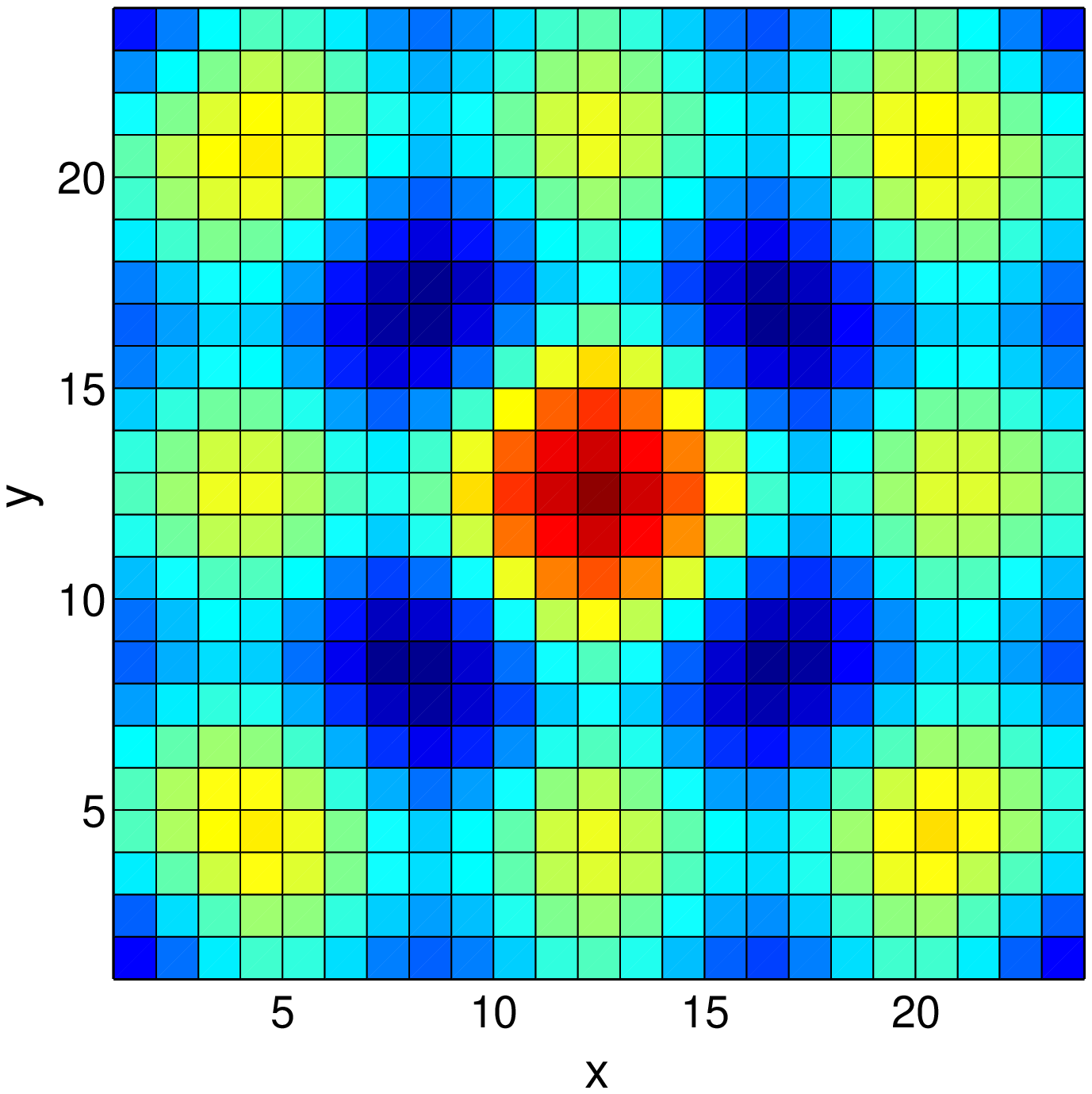,height=3.5cm,width=3.5cm,angle=0}}
\centerline{\psfig{figure=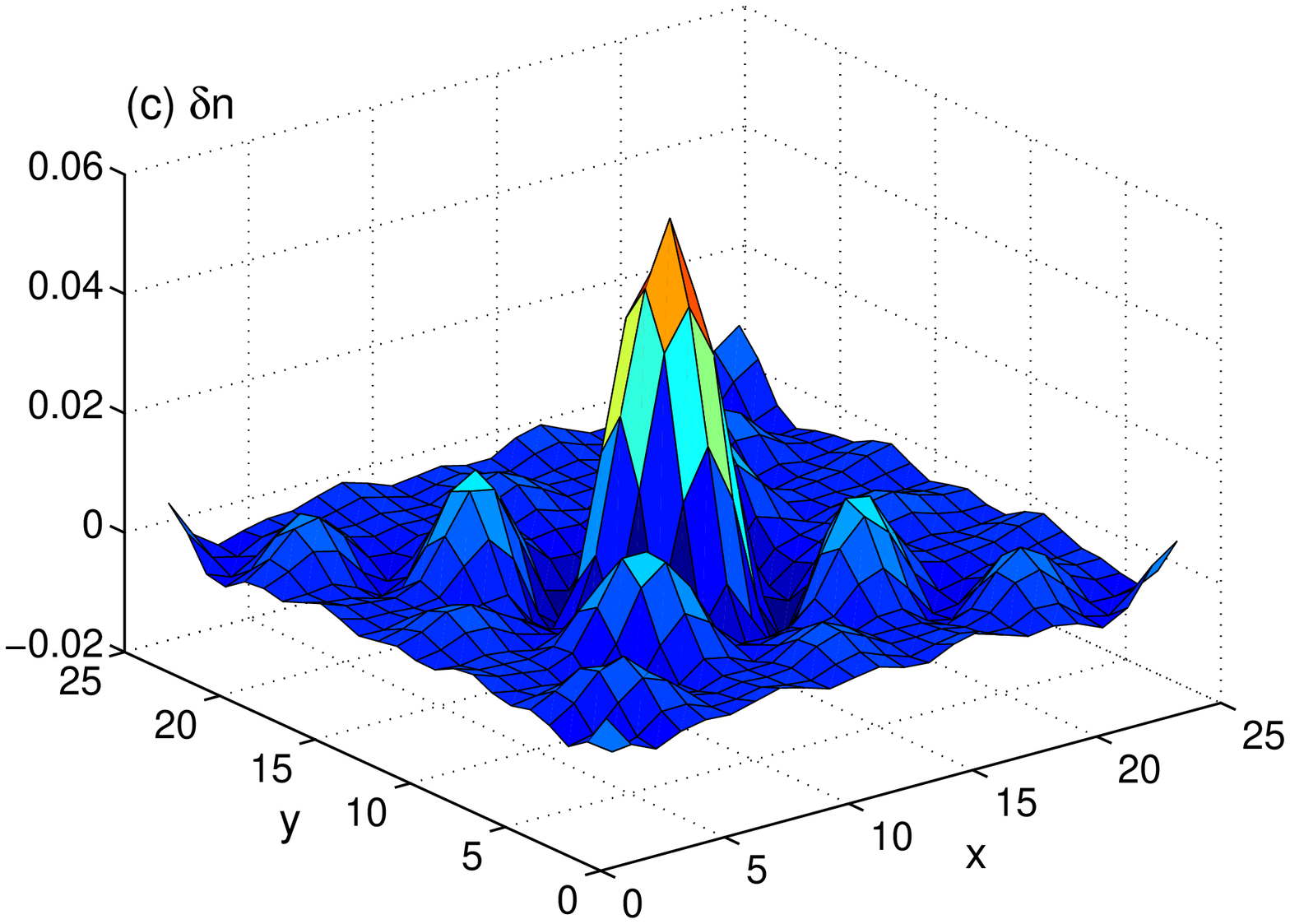,height=3.5cm,width=4.3cm,angle=0}
\psfig{figure=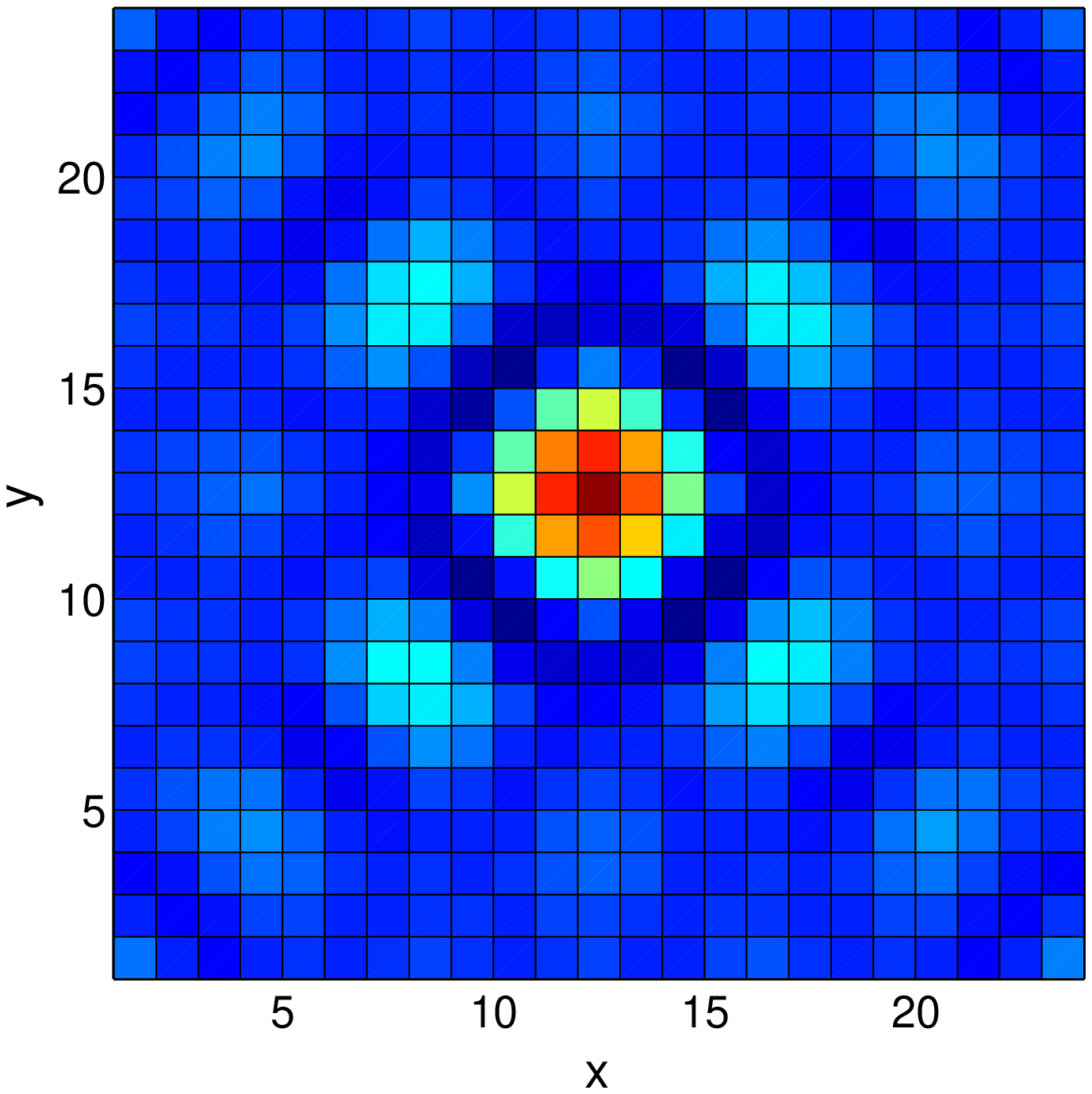,height=3.5cm,width=3.5cm,angle=0}}
\caption[*]{The three-dimensional (left column) and contour (right
column) display of the amplitude distribution of the $d$-wave SC
order parameter $\vert \Delta_{d}\vert$ (a), the staggered
magnetization $M_{s}$ (b), and the electron density $\delta
n=\sum_{\sigma}n_{i\sigma}-n_{f}$ (c) around one vortex located at
the center of an area of $24\times 24$ sites. The size of the
whole magnetic cell is $48\times 24$. Parameter values:
$t^{\prime}=-0.2$, $U=2.5$, $V=1$, and $n_f=0.84$. }
\label{FIG:VORTEX}
\end{figure}

\begin{figure}
\centerline{\psfig{figure=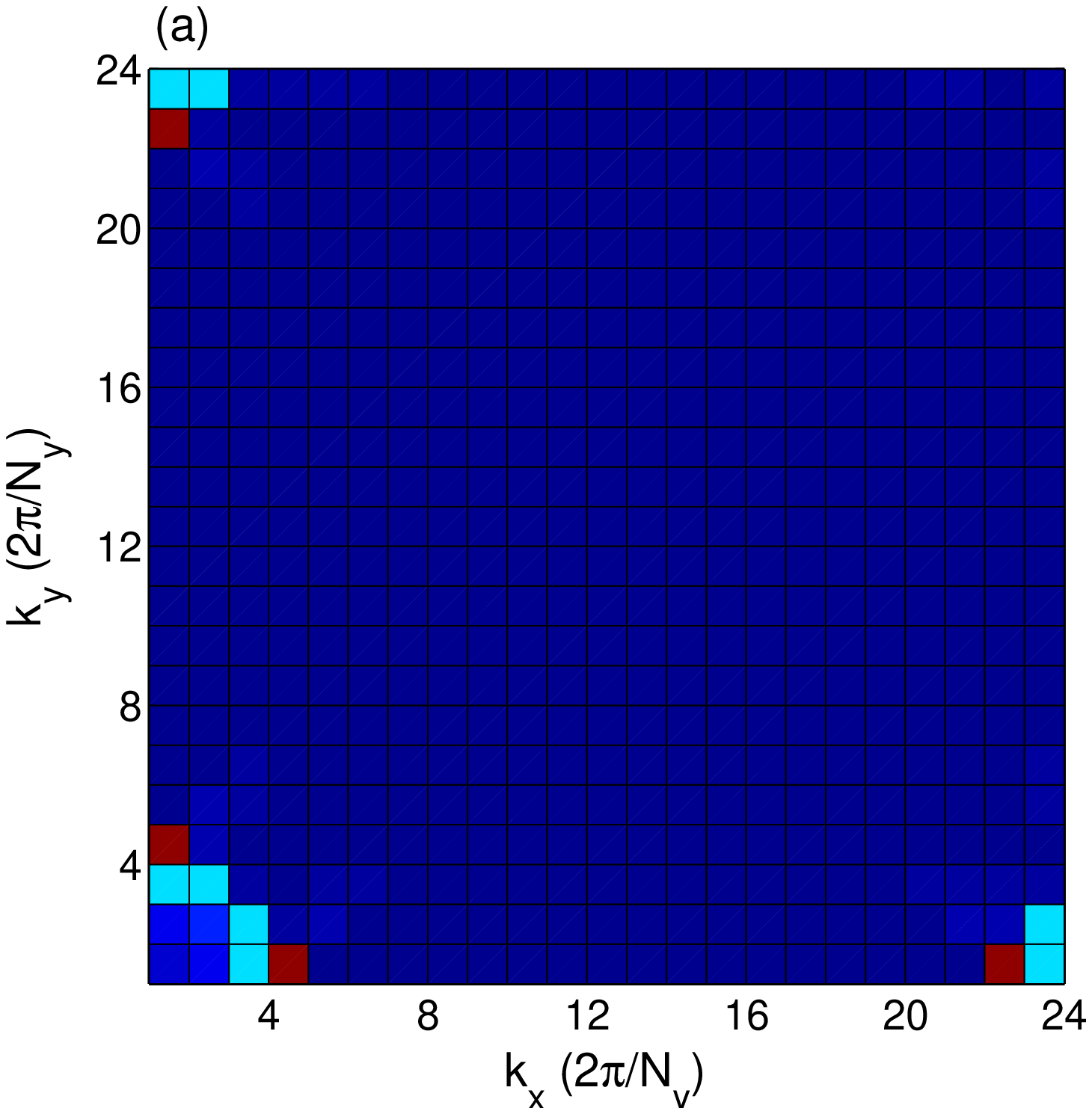,height=4cm,width=4cm,angle=0}
\psfig{figure=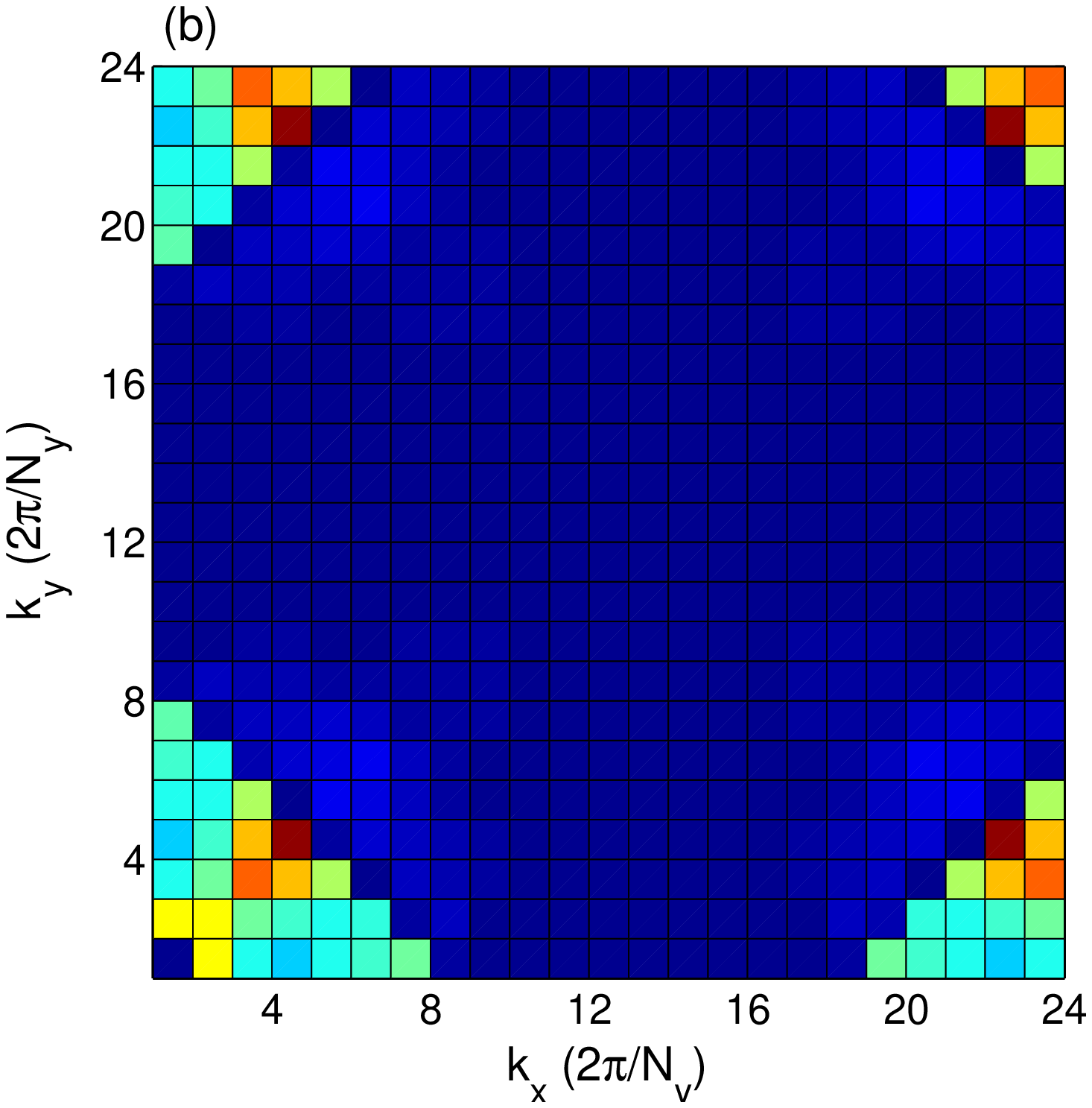,height=4cm,width=4cm,angle=0}}
\caption[*]{The Fourier transform of the spatial modulation of the
spin density $M_{s}$ (a) and the charge density $\delta n$ (b)
around the vortex. Parameter values are the same as in
Fig.~\ref{FIG:VORTEX}. } \label{FIG:VORTEX_FFT}
\end{figure}

{\em Field-induced SDW and CDW}.  When an external magnetic field
is applied perpendicular to the 2D Cu-O plane,
$\mathbf{H}=H\hat{\mathbf{z}}$ ($H_{c1}\ll H\ll H_{c2}$), an
Abrikosov vortex state is formed. As the vortex core is
approached, the DSC order parameter vanishes topologically due to
the circulating supercurrent surrounding the core. It is assumed
that  the superconductor is in the extreme type-II limit where the
Ginzburg-Landau parameter $\kappa=\lambda_0/\xi_0$ goes to
infinity so that the screening effect from the supercurrent is
negligible. We choose a Landau gauge to write the vector potential
as ${\bf A}=(-H y,0,0)$, where $y$ is the $y$-component of the
position vector {\bf r}, so that the Peierls phase factor
$\varphi_{ij}$ is uniquely determined. By taking the strength of
magnetic field, $H=2\Phi_0/N_x N_y$, such that the flux enclosed
by each unit cell is twice $\Phi_0$, we solve self-consistently
the BdG equation~(\ref{EQ:BdG}) with the aid of the magnetic Bloch
theorem as given by Eq.~(3) in Ref.~\cite{Zhu01}. For the
calculation, we consider the magnetic unit cell of size $N_x\times
N_y=48\times 24$, which gives a square vortex lattice. The
numerics shows that each unit cell accommodates two
superconducting vortices each carrying a flux quantum $\Phi_0$,
which conforms to the above prescription for the magnetic field
strength. The unit cell is equally partitioned between these two
vortices, each located at the center of area $\frac{N_x}{2} \times
N_y$ sites. Typical results on the structure around one vortex
core are displayed in Fig.~\ref{FIG:VORTEX}, where the left column
is three-dimension plots and the right column is contour plots. As
shown in Fig.~\ref{FIG:VORTEX}(a), the DSC order parameter
vanishes at the core center (dark-blue site) and approaches its
zero-field value, which is about $\Delta_0=0.08$ for the chosen
parameter values, away from the core center. Notice that the DSC
order parameter is not uniform beyond the distance $\xi_0$ away
from the core center. Instead, it is weakly modulated. The maximum
modulation amplitude is less than $0.05 \Delta_0$. This modulation
is closely related to the appearance of the field-induced SDW, as
will be discussed immediately. Fig.~\ref{FIG:VORTEX}(b) displays
the spatial distribution of the staggered magnetization of the
local SDW order defined as
$M_{s}=(-1)^{i}(n_{i\uparrow}-n_{i\downarrow})$. Clearly, the
maximum strength of $M_{s}$ is pinned at the vortex core center.
This AF SDW order exhibits a modulation pattern with the satellite
peaks (yellow spots) and valleys (dark-blue spots) regularly
spaced throughout the unit cell along the Cu-O bond directions.
The modulation pattern of the SDW implies a much longer magnetic
correlation length as compared to $\xi_0$, which is consistent
with the INS measurements~\cite{Lake01}. The appearance of the SDW
order around the vortex core also strongly affects the electron
density $n_{i}=\sum_{\sigma} n_{i\sigma}$. As shown in
Fig.~\ref{FIG:VORTEX}(c), at the vortex core center, where the SDW
amplitude reaches the global maximum, the electron density is
strongly enhanced. In addition, the charge density also exhibits
regular modulation. By comparing the spatial distribution of the
DSC order parameter, the field-induced SDW as well as the
associated CDW, one finds that as long as the absolute amplitude
($\vert M_s\vert$) of SDW reaches a local maximum (both yellow and
dark-blue spots in Fig.~\ref{FIG:VORTEX}(b)), the associated CDW
also reaches a local maximum (green spots in
Fig.~\ref{FIG:VORTEX}(c)), that is, a local minimum in the hole
density,  while the DSC order parameter has a local minimum
(shallow red spots in Fig.~\ref{FIG:VORTEX}(a)). The one to one
correspondence between  the SDW and associated CDW orderings at
zero field, was discussed in the context of
stripes~\cite{Poilblanc89,Zaanen89,Emery90,Zachar98,Martin00}. To
make a closer inspection of the periodicity of the SDW and
associated CDW modulation, we perform a Fourier transform of these
two quantities. As shown in Fig.~\ref{FIG:VORTEX_FFT}(a), the
strongest spectral intensity of the SDW modulation occurs at the
wave vectors $\mathbf{k}=2\pi(\frac{3}{24},0)$ and
$\mathbf{k}=2\pi(0,\frac{3}{24})$, which gives unambiguously the
period of $8a_0$ along the Cu-O bond directions.  To verify this
point, we have performed a calculation on the unit cell of
different size $N_x\times N_y=52\times 26$, and found that the
period of the SDW modulation remains $8a_0$, indicating
convincingly that the periodicity of the field-induced SDW
modulation is an intrinsic property. Therefore, our result
explains very well the INS measured value of the SDW
modulation~\cite{Lake01}. We notice that, in the optimal doping
regime, the circulating supercurrent around the vortex, which
favors the isotropy between $x$ and $y$ directions, overcomes the
one-dimensionality of stripes. In the underdoping regime, the
stripe behavior becomes dominant~\cite{Ichioka01}.

\begin{figure}
\centerline{\psfig{figure=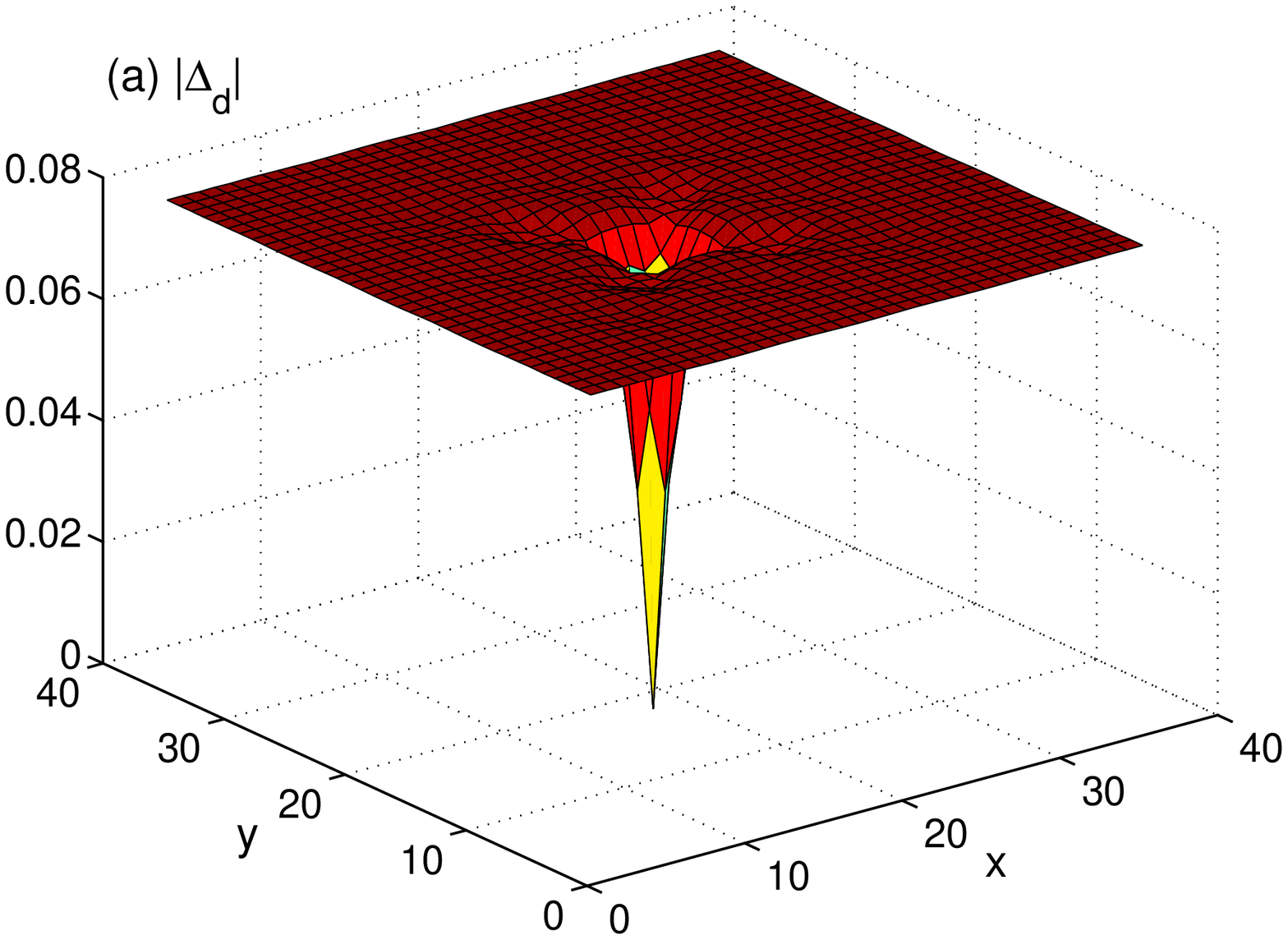,height=3.5cm,width=4.3cm,angle=0}
\psfig{figure=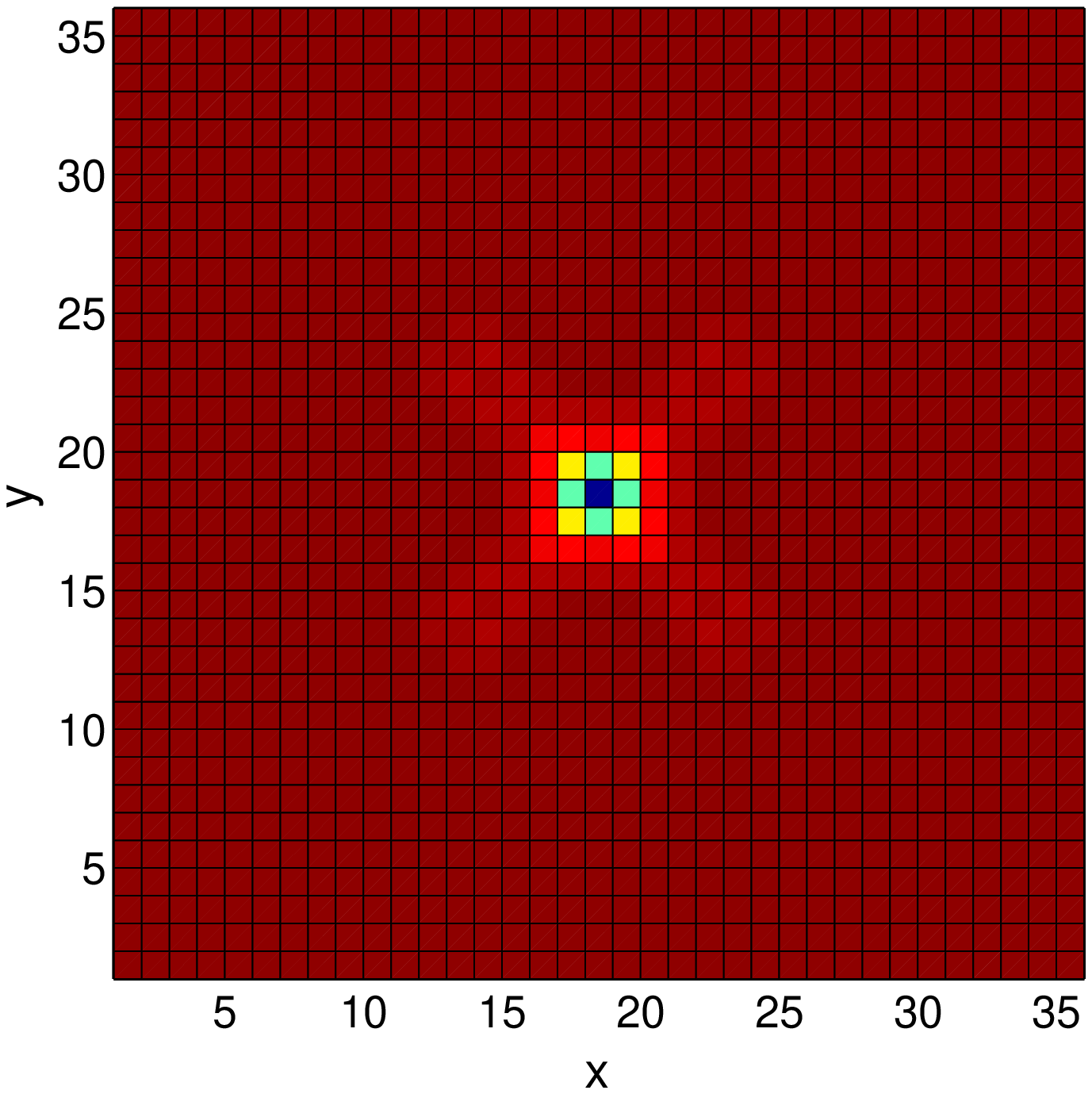,height=3.5cm,width=3.5cm,angle=0}}
\centerline{\psfig{figure=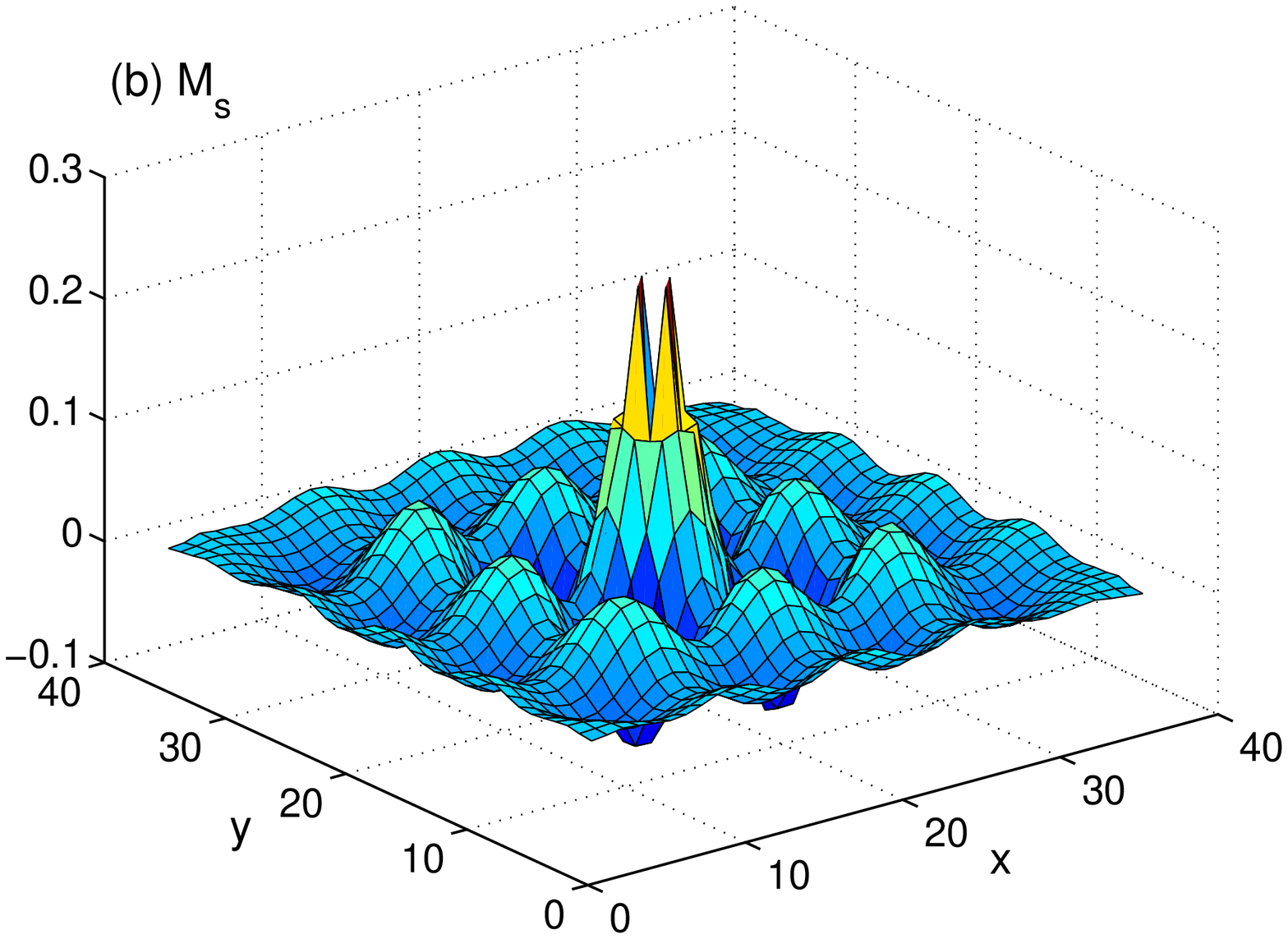,height=3.5cm,width=4.3cm,angle=0}
\psfig{figure=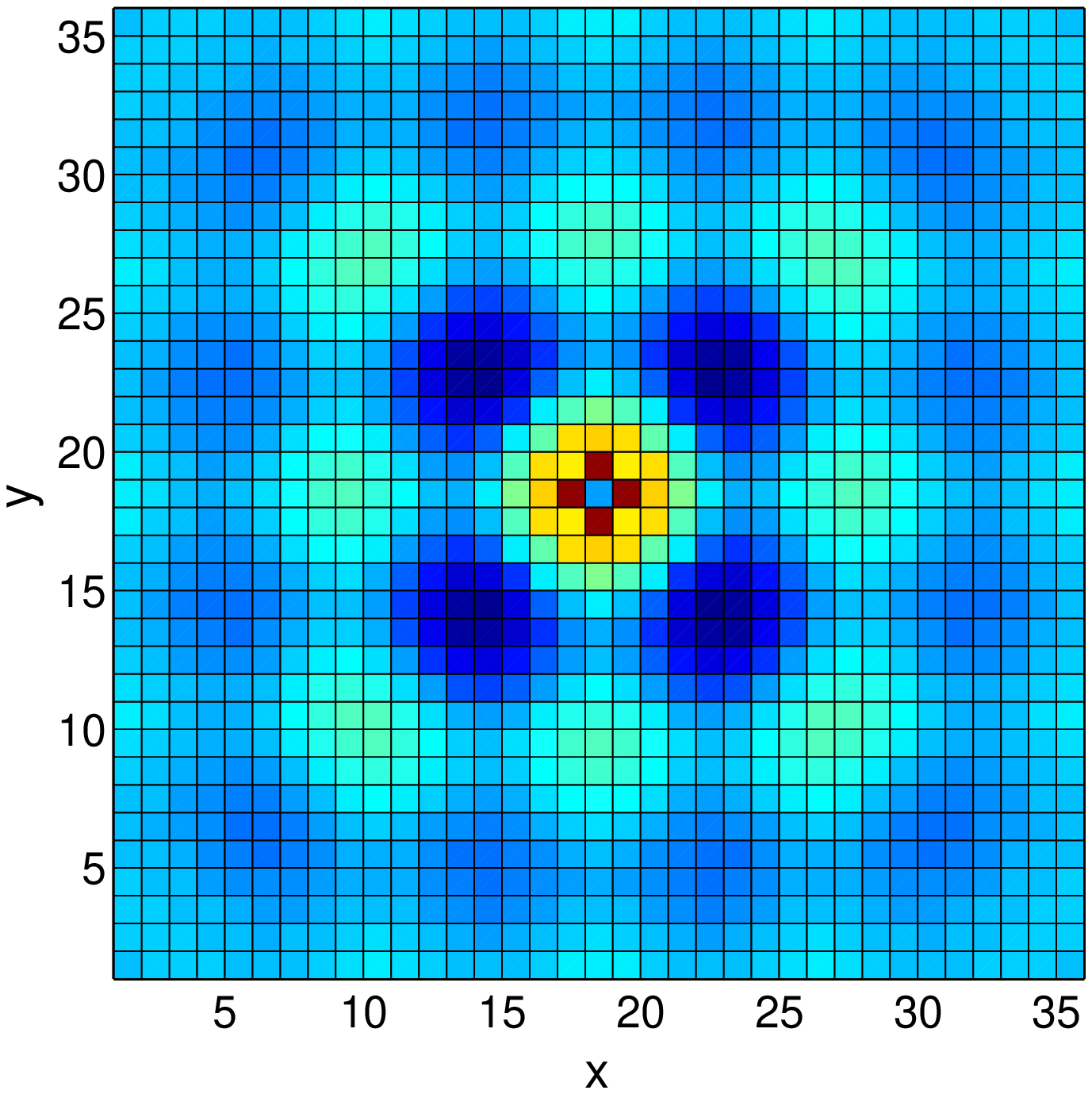,height=3.5cm,width=3.5cm,angle=0}}
\centerline{\psfig{figure=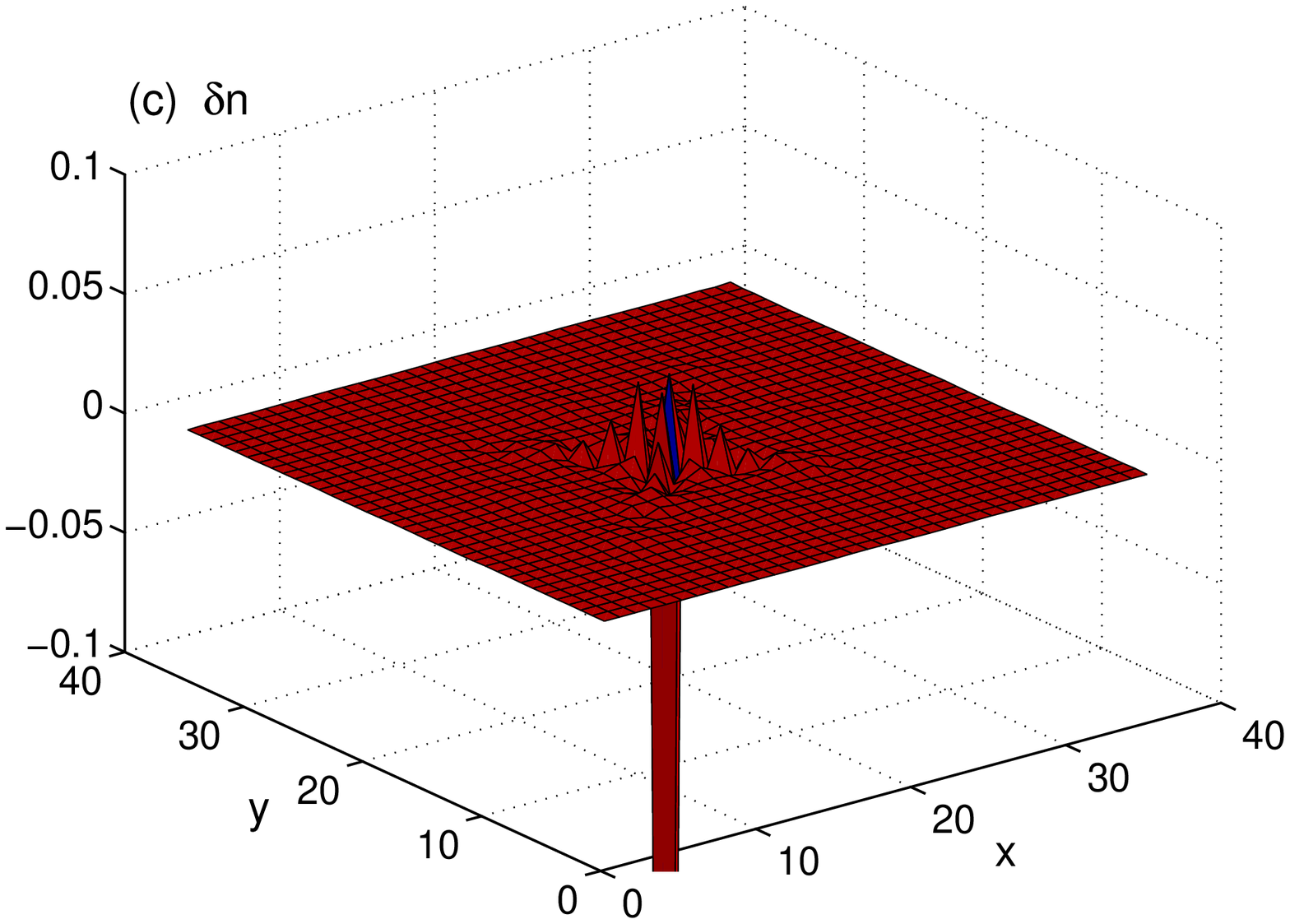,height=3.5cm,width=4.3cm,angle=0}
\psfig{figure=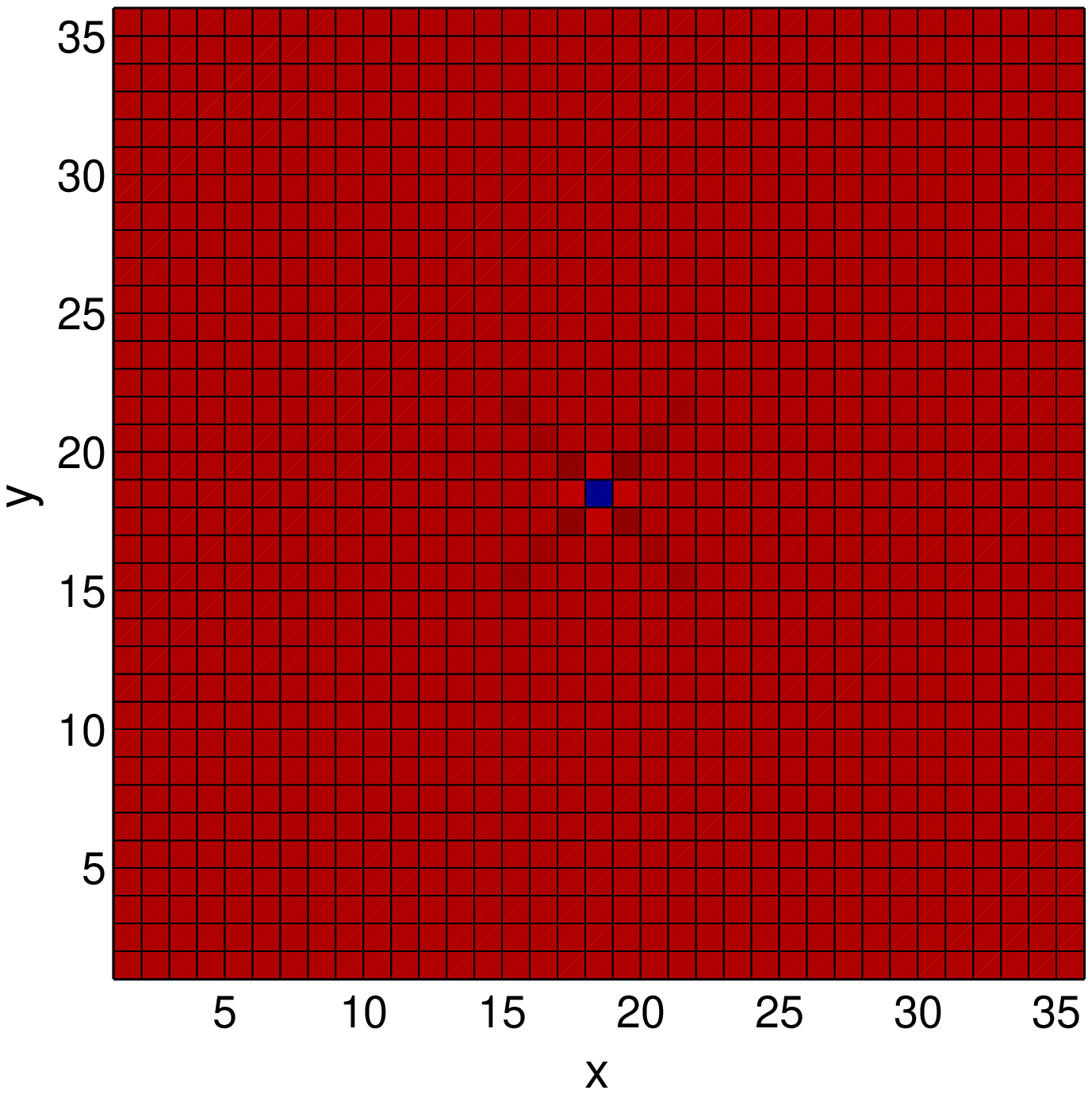,height=3.5cm,width=3.5cm,angle=0}}
\caption[*]{The three-dimensional (left column) and contour (right
column) display of the amplitude distribution of the $d$-wave SC
order parameter $\vert \Delta_{d}\vert$ (a), the staggered
magnetization $M_{s}$ (b), and the electron density $\delta
n=n_{i}-n_{f}$ (c) around a nonmagnetic unitary impurity
($\epsilon_0=100$) located at the center of the unit cell of
$36\times 36$ sites. Parameter values are the same as in
Fig.~\ref{FIG:VORTEX}.} \label{FIG:IMP}
\end{figure}

\begin{figure}
\centerline{\psfig{figure=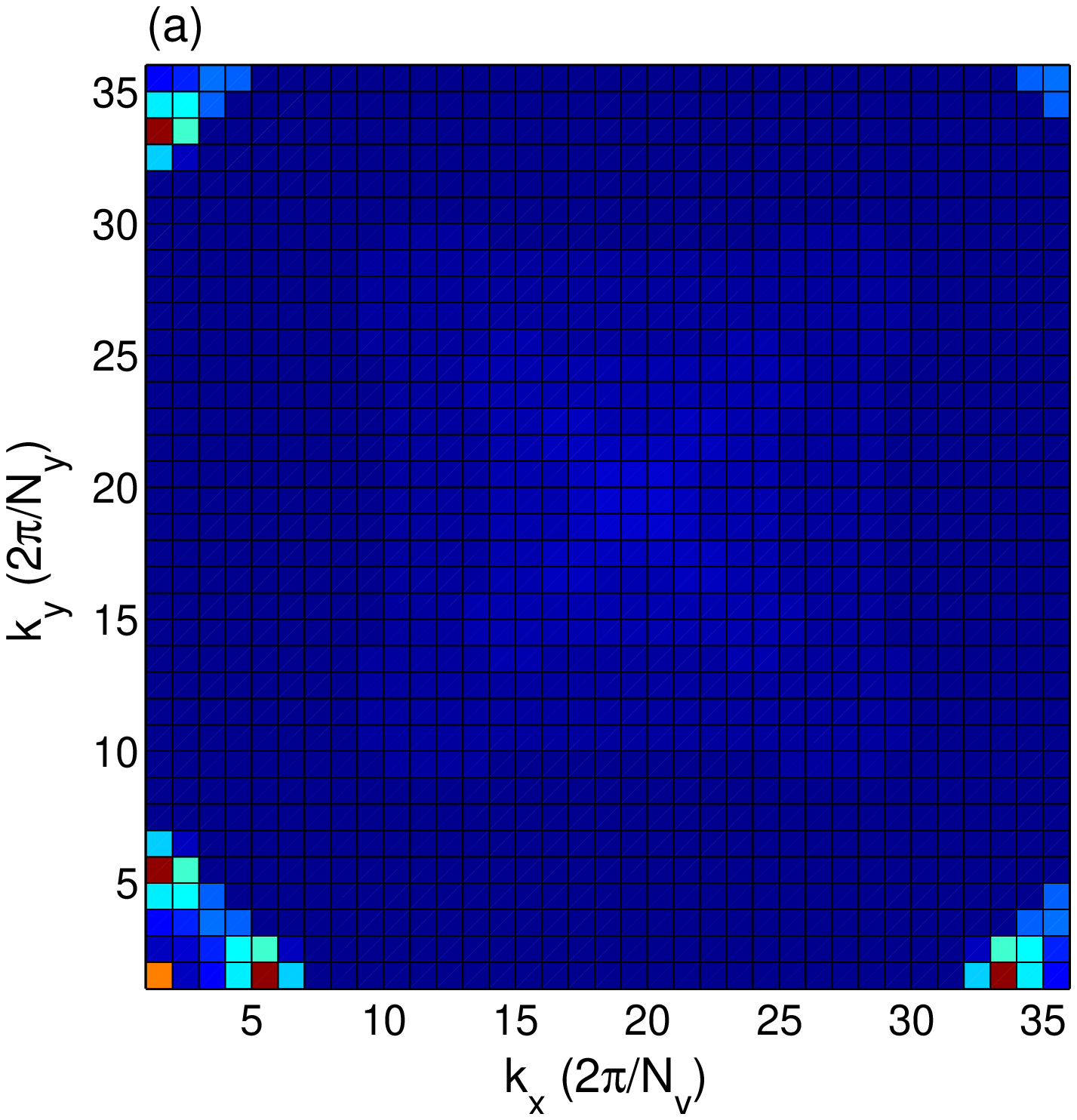,height=4cm,width=4cm,angle=0}
\psfig{figure=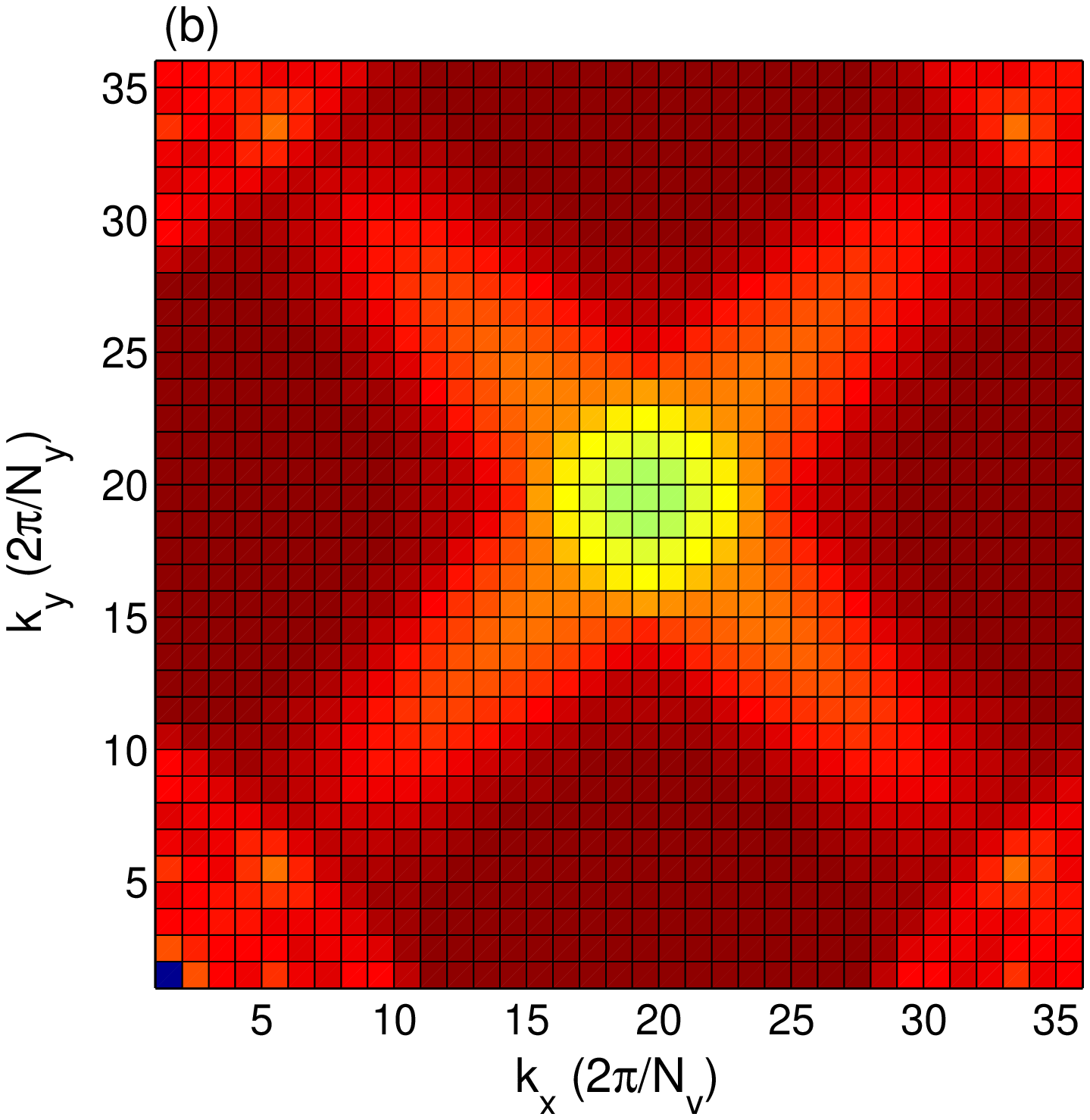,height=4cm,width=4cm,angle=0}}
\caption[*]{The Fourier transform of the spatial modulation of the
spin density $M_{s}$ (a) and the charge density $\delta n$ (b),
around a nonmagnetic unitary impurity. Parameter values are the
same as in Fig.~\ref{FIG:IMP}. } \label{FIG:IMP_FFT}
\end{figure}

{\em Impurity-induced SDW and CDW}. The $d$-wave superconductivity
can also be suppressed locally by a strong nonmagnetic impurity.
The effects of impurities on superconductors have been of
theoretical and experimental interest even in its own right for a
long time. In view of the recent observations of the AF magnetism
around the vortex core, it is important to see whether a similar
magnetic structure can also be induced around an impurity. We
perform the numerical calculation on a unit cell of $N_x\times N_y
=36 \times 36$ sites. The single-site potential is taken to be
$\epsilon_0=100$ at the unitary impurity and zero on other sites.
Since $\varphi_{ij}=0$ in this case, the conventional Bloch
theorem is used. The results on the spatial distribution of the
three orderings is displayed in Fig.~\ref{FIG:IMP}. The DSC order
parameter is depressed dramatically at the impurity site and
approaches the bulk value at the scale $\xi_0$. Out of this range,
almost no modulation of the DSC order parameter is seen. The
induced staggered moment of the SDW is zero at the impurity site
and has maxima on the four nearest neighbor Cu sites of the
impurity. Away from the impurity, the induced SDW shows a
modulation very similar to the vortex case, although with a much
weaker amplitude. However, the electron density, which is zero at
the impurity, only exhibits a Friedel-like oscillation within a
limited range around the impurity. The Fourier transform, as shown
in Fig.~\ref{FIG:IMP_FFT}, gives a clear picture: The main
modulation period of the SDW is $8a_0$; the CDW is modulated with
a period of $2a_0$, which is roughly the Fermi wave length.
Therefore, the correspondence between the SDW and CDW is absent in
this impurity case. The reason lies in the fact that the CDW is
very sensitive to the strong impurity scattering. This result is
consistent with the STM measurements in BSCCO~\cite{Pan00b} that
show no evidence for CDW modulation around the impurities.
However, SDW appears to be a robust feature induced by the
impurity, which should be observable in neutron scattering
experiments.

Nuclear magnetic resonance (NMR) measurements have shown that when
a Cu$^{2+}$ in the Cu-O plane is substituted by a strong
nonmagnetic impurity, such as Zn$^{2+}$, an effective magnetic
moment can be induced on the Cu sites around the impurity
site~\cite{Alloul91,Ishida96,Julien00,Bobroff01}. More recent NMR
measurements~\cite{Bobroff01} show for the first time that near
optimal doping, the Kondo screening effect observed above the
superconducting transition temperature, is strongly reduced in the
superconducting state. This indicates the stabilization of the
magnetic moments. Our result of the strongest staggered magnetic
moments on the four nearest-neighbor Cu sites of the impurity is
consistent with the above experiment~\cite{Bobroff01}. This leads
us to speculate that the magnetism around vortices and around
impurities may share a common origin.

In conclusion, we have studied the magnetism around vortices and
nonmagnetic unitary impurities. In the case vortices, the
experimentally observed $8a_0$ period of the SDW modulation is
explained for the first time based on a microscopic model. The
correspondence between the SDW and CDW modulations has also been
established.  Around the unitary impurity, we have also shown the
existence of the SDW modulation with a period of $8a_0$. The
existence of such modulation can be tested by neutron scattering
experiments.

{\bf Acknowledgments}: We wish to thank A.V. Balatsky, J.C. Davis,
B. Khaykovich, G. Ortiz, S. Sachdev, C.S. Ting  for useful
discussions. This work was supported by the US Department of
Energy through the Los Alamos National Laboratory.


\begin{thebibliography}{99}

\bibitem{Lake01} B. Lake {\em et al.}, Science {\bf 291}, 1759
(2001).

\bibitem{Lake02} B. Lake {\em et al.}, Nature {\bf 415}, 299
(2002).

\bibitem{Khaykovich01} B. Khaykovich {\em et al.},
cond-mat/0112505.

\bibitem{Mitrovic01} V. F. Mitrovi\'{c} {\em et al.}, Nature {\bf
413}, 501 (2001).

\bibitem{Hoffman02} J. E. Hoffman {\em et al.}, Science {\bf 295},
466 (2002).

\bibitem{Demler01} E. Demler, S. Sachdev, and Y. Zhang, Phys. Rev.
Lett. {\bf 87}, 167004 (2001).

\bibitem{Zhang01} Y. Zhang, E. Demler, and S. Sachdev,
cond-mat/0112343.

\bibitem{Zhang97} S.-C. Zhang, Science {\bf 275}, 1089 (1997).

\bibitem{Arovas97} D. P. Arovas {\em et al.}, Phys. Rev. Lett. {\bf 79},
2871 (1997).

\bibitem{Hu01} J.-P. Hu and S.-C. Zhang, cond-mat/0108273.

\bibitem{Poilblanc89} D. Poilblanc and T. M. Rice, Phys. Rev. B
{\bf 39}, 9749 (1989).

\bibitem{Zaanen89} J. Zaanen and O. Gunnarson, Phys. Rev. B {\bf
40}, 7391 (1989).

\bibitem{Emery90} V. J. Emery, S. Kivelson, and H. Lin, Phys. Rev.
Lett. {\bf 64}, 475 (1990); V. J. Emery and S. Kivelson, Physica
(Amsterdam) {\bf 209C}, 597 (1993).

\bibitem{Zachar98} O. Zachar, S. A. Kivelson, and V. J. Emery,
Phys. Rev. B {\bf 57}, 1422 (1998).

\bibitem{Zhu01} Jian-Xin Zhu and C. S. Ting, Phys. Rev. Lett. {\bf
87}, 147002 (2001).



\bibitem{Martin00} I. Martin {\em et al.}, Int. J. Mod. Phys.
B {\bf 14}, 3567 (2000).

\bibitem{Maggio95} I. Maggio-Aprile {\em et al.}, Phys. Rev. Lett.
{\bf 75}, 2754 (1995).

\bibitem{Pan00} S. H. Pan {\em et al.}, Phys. Rev. Lett. {\bf 85}, 1536
(2000).

\bibitem{Hoogenboom01} B. W. Hoogenboom {\em at al.}, Phys. Rev. Lett. {\bf 87},
267001 (2001); Ch. Renner {\em et al.}, Phys. Rev. Lett. {\bf 80},
3606 (1998).

\bibitem{Ichioka01} M. Ichioka, M. Takigawa, and K. Machida, J.
Phys. Soc. Jpn. {\bf 70}, 33 (2001).

\bibitem{Pan00b} S. H. Pan {\em et al.}, Nature {\bf 403}, 746
(2000).


\bibitem{Alloul91} H. Alloul {\em et al.}, Phys. Rev. Lett. {\bf
67}, 3140 (1991); A. V. Mahajan {\em et al.}, {\em ibid.} {\bf
72}, 3100 (1994); J. Bobroff {\em et al.}, {\em ibid.} {\bf 83},
4381 (1999).

\bibitem{Ishida96} K. Ishida {\em et al.}, Phys. Rev. Lett. {\bf
76}, 513 (1996).

\bibitem{Julien00} M.-H. Julien {\em et al.}, Phys. Rev. Lett.
{\bf 84}, 3422 (2000).

\bibitem{Bobroff01} J. Bobroff {\em et al.}, Phys. Rev. Lett. {\bf
86}, 4116 (2001).


\end{thebibliography}
\end{document}